\documentclass[aps,prb,twocolumn,preprintnumbers,amsmath,amssymb,nofootinbib,floatfix]{revtex4}
\pdfoutput=1
\usepackage{graphicx}		
\usepackage{dcolumn}		
\usepackage{bm}			

\begin{document}

\title{Electronic structure of copper intercalated transition metal dichalcogenides: First-principles calculations}
\author{R. A. Jishi}
\author{H. M. Alyahyaei}
\affiliation{ Department of Physics, California State University, Los Angeles, California 90032}

\date{\today}

\begin{abstract}
We report first principles calculations, within density functional theory, of copper intercalated titanium diselenides, Cu$_x$TiSe$_2$, for values of $x$ ranging from 0 to 0.11.  The effect of intercalation on the energy bands and density of states of the host material is studied in order to better understand the cause of the superconductivity that was recently observed in these structures.  We find that charge transfer from the copper atoms to the metal dichalcogenide host layers causes a gradual reduction in the number of holes in the otherwise semi-metallic pristine TiSe$_2$, thus supressing the charge density wave transition at low temperatures, and a corresponding increase in the density of states at the Fermi level.  These effects are probably what drive the superconducting transition in the intercalated systems.
\end{abstract}

\maketitle

\section{\label{sec:introduction}Introduction}
Transition metal dichalcogenides (TMDCs) are quasi-two-dimensional, highly anisotropic layered compounds that are of great interest in basic research as well as in applications in such areas as lubrication, catalysis, electrochemical photocells, and novel battery systems~\cite{Fried_Yoffe:1987}.  Recently, it became possible to synthesize other forms of TMDCs such as balls and nanotubes~\cite{Tenne_Margolis:1992,Frey_Elani:1998,Remskar_Mrzel:2001}.  TMDCs generally have the formula MX$_2$, where M is a transition metal, such as vanadium, titanium, tantalum, molybdenum, or others, and X is a chalcogen atom, such as sulfur, selenium, or tellurium.  Each layer consists of a metal sheet sandwitched between two chalcogen sheets.  Whereas within each layer the bonds connecting a metal atom in one sheet to the chalcogen atoms in surrounding sheets are strong covalent bonds, adjacent layers are only weakly coupled through the van der Waal's interaction.  Two types of MX$_2$ sandwitches occur depending on the coordination of the transition metal atom by the chalcogens.  In 1T-MX$_2$ the coordination is octahedral whereas in 2H-MX$_2$ it is trigonal prismatic.

The weak coupling between adjacent layers in TMDCs makes it possible to introduce foreign species into the region between these layers, a region called the van der Waal's gap.  This process, known as intercalation, provides the ability to tune the electronic properties of the system by controlling the type and concentration of the intercalant.  This is a general property of layered materials, and just like graphite, which has been intercalated with many kinds of atoms and molecules,~\cite{Dresselhaus:1981} TMDCs have been intercalated with alkali atoms, transition metal atoms, and organic and inorganic molecules~\cite{Fried_Yoffe:1987}.

Among the most important properties of TMDCs is that some of them undergo a phase transition to a charge density wave (CDW) state as the temperature is lowered~\cite{Wilson_DiSalvo:1975}. 1T-TiSe$_2$ is one of the first compounds where the CDW state was observed at a temperature of 200K~\cite{Wilson_DiSalvo:1975,Wilson_Yoffe:1969,DiSalvo_Moncton:1976}.  The state is associated with the formation of a commensurate (2x2x2) superlattice.  It is not completely clear what mechanism drives this phase transition.  Possible explanations are offered in terms of an indirect Jahn-Teller effect,~\cite{Kidd_Miller:2002,Snow_Karpus:2003} or by an exciton condensation mechanism~\cite{Kohn:1967,Halperin_Rice:1968}.

Recent angle resolved photoemission spectroscopy (ARPES) studies have not resolved whether 1T-TiSe$_2$ is a semimetal with low electron and hole concentraion or a narrow-gap semiconductor in either the normal or the CDW state~\cite{Zhao_Ou:arXiv,Negishi_Negishi:2006,Cui_Negishi:2006}.  Optical measurements, on the other hand, reveal that TiSe$_2$ is a semimetal with low carrier density at temperatures above as well as below the CDW transition~\cite{Li_Hu:2007}.  This is consistent with band structure calculations~\cite{Zunger_Freeman:1978} which reveal the presence of an electron pocket, around the L point in the Brillouin zone, derived from Ti 3d states, and a hole pocket around the $\Gamma$ point derived from Se 4p states.

Recently it was reported~\cite{Morosan_Zandbergen:2006} that upon controlled intercalation of TiSe$_2$ with Cu to yield Cu$_x$TiSe$_2$ the CDW transition is continually suppressed, making way for a new superconducting state that emerges near $x=0.04$, with a maximum transition temperature T$_c$ of 4.15K ocurring at $x=0.08$;  further increase in $x$ results in a drop of T$_c$.  It should be noted that while superconductivity was previously observed in 2H structures, such as 2H-NbSe$_2$, 2H-TaSe$_2$, 2H-NbS$_2$, and 2H-TaS$_2$,~\cite{CastroNeto:2001} the copper intercalated TiSe$_2$ is the first superconducting 1T-structured TMDC.  As Cu atoms are intercalated, they move into the van der Waal's gap separating adjacent TiSe$_2$ layers.  With a single electron in the outer 4s shell, electronic charge transfer proceeds from Cu to the TiSe$_2$ layers, causing an increase in the number of electrons in the Ti 3d bands and a reduction in the number of holes in the Se 4p bands.  As a result of this, the effective screening of the electron-hole Coulumb interaction is enhanced, causing a drop in the CDW transition temperature as the intercalant concentration increases, and at $x=0.06$ the transition is completelly supressed.  The superconducting transition, on the other hand, emerges at $x=0.04$.  The Cu$_x$TiSe$_2$ phase diagram resembles the corresponding diagram in cuprate and heavy fermion superconductors, except that there it is the antiferromagnetic spin order that competes with superconductivity, whereas in Cu$_x$TiSe$_2$ it is the charge order~\cite{Dagotto:2005}.

Recently, calculations of density of states (DOS) and the electron phonon coupling, using linear muffin-tin orbital (LMTO) method within the local density approximation, were reported for the case $x=0.08$ by Jeong and Jarlborg~\cite{Jeong_Jarlborg:2007}.  The authors consider a 2x2x3 supercell, Cu(TiSe$_2$)$_{12}$, and conclude that charge transfer from copper raises the electronic density of states at the Fermi energy.  They also find a strong dependence on pressure of the DOS, and consequently the superconducting transition temperature.

In this paper we report band structure calculations for Cu$_x$TiSe$_2$ for values of $x$ ranging from $0.0$ to $0.11$ in an attempt to study the effect of copper intercalation on the electronic structure of TiSe$_2$.  In particular, we calculate the energy bands of TiSe$_2$ and the compounds Cu$_{1/16}$TiSe$_2$, Cu$_{1/13}$TiSe$_2$, and Cu$_{1/9}$TiSe$_2$.  In Section~\ref{sec:method} we discuss the computational method and some relevant details, and in Section~\ref{sec:results_and_discussion} we discuss the results and conclusions.

\section{\label{sec:method}Method}
The first-principles calculations presented in this work were performed using the all-electron full potential linear augmented plane wave plus local orbitals (FP-LAPW+lo) method as implemented in WEN2K code~\cite{Blaha_Schwarz:2001}.  In this method the core states are treated in a fully relativistic way but the valence states are treated at a scalar relativistic level.  The exchange-correlation potential was calculated using the generalized gradient approximation (GGA) as proposed by Pedrew, Burke, and Ernzerhof~\cite{Perdew_Burke:1996}.

In the calculations on 1T-TiSe$_2$ the experimentally measured lattice constants ($a$ = $b$ = 3.534\AA, $c = 6.008$\AA) are employed~\cite{Riekel:1976}. The distance betwen the Ti plane and either of the adjacent Se planes is $d=zc$, where the value of $z$ is obtained by minimizing the forces on the atoms to below 2.0 mRy/Bohr.  In calculations on the intercalated compounds Cu$_x$TiSe$_2$ the lattice constants are taken to coincide with the experimental values~\cite{Morosan_Zandbergen:2006}, but the internal structure of the unit cell is allowed to relax, subject only to the constraint that the space group remains unchanged; again the force on each atom is reduced to below 2.0 mRy/Bohr.  For the compound Cu$_{1/16}$TiSe$_2$ the structure is taken to be that of TiSe$_2$ 4x4 superlattice in the $a$-$b$ plane, with one Cu atom placed midway between one Ti atom in one layer and the nearest Ti atom in the adjacent layer; the unit cell is then hexagonal with $a'=b'=4a$.  Accordingly, the unit cell has 1 Cu atom, 16 Ti atoms, and 32 Se atoms, and if one Ti atom is placed at the origin (0,0,0), then the Cu atom is placed at (0,0,$c'$/2), where $c'$ is the lattice parameter, in the direction normal to the $a$-$b$ plane, in the intercalated systems.  A similar situation holds for Cu$_{1/9}$TiSe$_2$ and Cu$_{1/13}$TiSe$_2$ where the corresponding superlattices in the $a$-$b$ plane are 3x3 and $\sqrt{13}$x$\sqrt{13}$, respectively.  The space group is P-3m1(\#164), the same as in TiSe$_2$, in all these structures, except in Cu$_{1/13}$TiSe$_2$, where it is P-3(\#147).  It should be noted here that the superlattice structure employed in these calculations is not necessarily the experimental lattice structure.  For example, for the case $x=0.0625$, we assume that the system is periodic with lattice constants in the $a$ and $b$ directions that are 4 times bigger than the corresponding values for $x=0$, whereas the experimental crystal structure is not reported and there may well be some disorder in the system such that the Cu atoms, residing between adjacent TMDC layers, do not form a perfectly periodic two-dimensional lattice.

For calculations in this work, the radii of the muffin-tin spheres for Cu, Ti, and Se atoms were taken to be 2.5a$_0$, 2.5a$_0$, and 2.25a$_0$ respectively, where a$_0$ is the Bohr radius.  The number of k-points in the irreducible part of the Brillouin zone (BZ) were 38, 41, 47, and 360 for Cu$_{1/16}$TiSe$_2$, Cu$_{1/13}$TiSe$_2$, Cu$_{1/9}$TiSe$_2$, and TiSe$_2$, respectively; these numbers were chosen so that the density of k-points in the whole BZ is roughly the same in the four cases.  For all structures considered in this work we set the parameter R$_{\text{MT}}$K$_{\text{MAX}}$=7, where R$_{\text{MT}}$ is the smallest muffin-tin radius and K$_{\text{MAX}}$ is a cutoff wave-vector.  The valence wavefunctions inside the muffin-tin spheres are expanded in terms of spherical harmonics up to $l_{\text{max}}$ = 10 while in the interstitial region they are expanded in plane waves with a wave-vector cutoff K$_{\text{MAX}}$, and the charge density is Fourier expanded up to G$_{\text{MAX}}$=14a$_0^{-1}$.  Convergence of the self consistent field calculations is attained with a total energy convergence tolerance of 0.1 mRy.

\begin{figure}
   \includegraphics[width=0.5\textwidth]{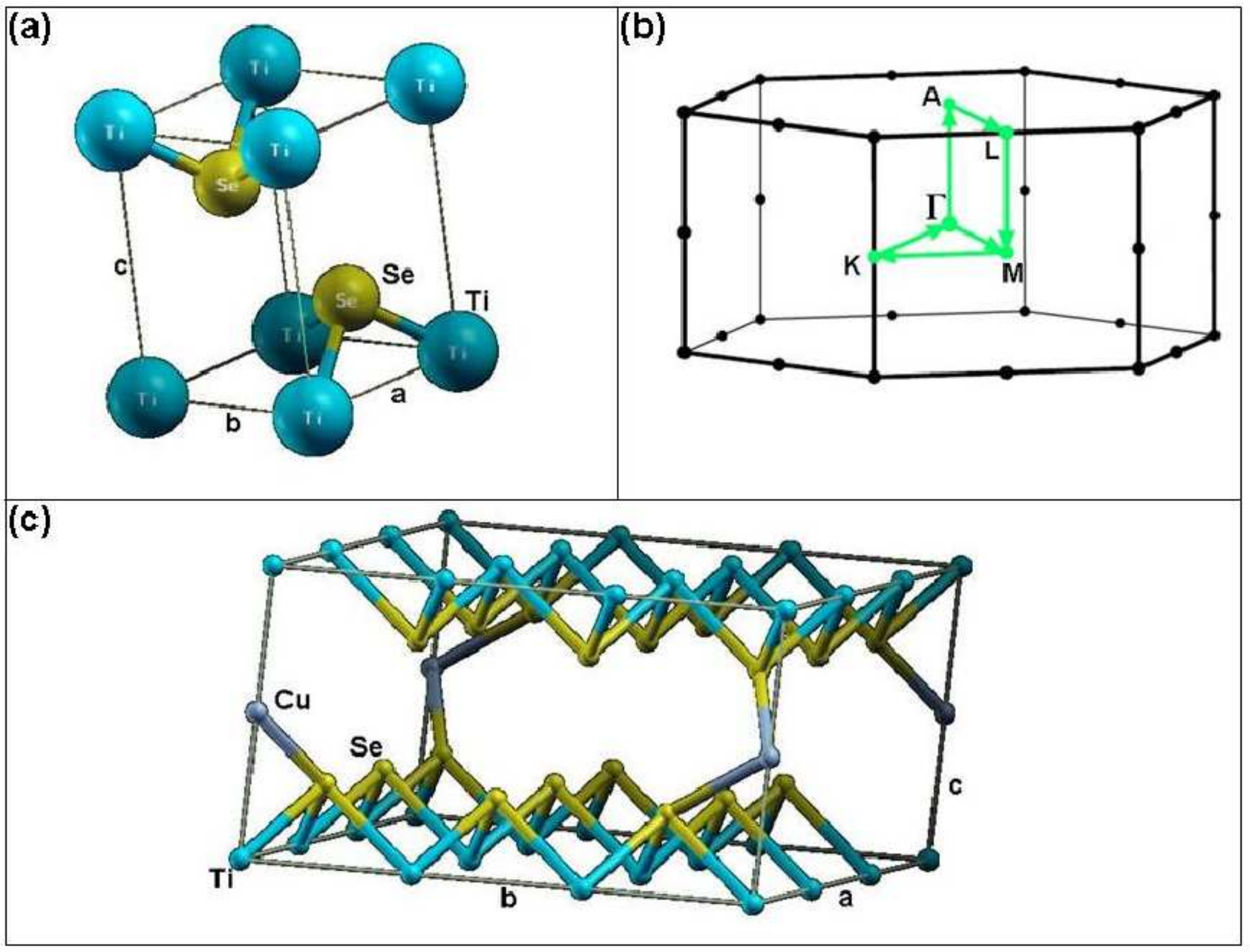}
   \caption{\label{fig:crystal}(color online)The unit cell of TiSe$_2$ is shown in (a), while the Brillouin zone, and the k-path along which the energy bands are calculated, are shown in (b).  In (c), the unit cell for Cu(TiSe$_2$)$_9$ is shown.}
\end{figure}

\begin{table}
	\begin{ruledtabular}
	\begin{tabular}{ l c | c c c | c c c }
		\multicolumn{3}{c}{\ } &  \multicolumn{2}{c}{$a$/\AA} & \multicolumn{2}{c}{$c$/\AA} & \ \\ \hline
		\multicolumn{3}{c}{\ } &  \multicolumn{2}{c}{10.641} & \multicolumn{2}{c}{6.044} & \  \\ \hline
		\multicolumn{2}{c|}{\ } & \multicolumn{3}{c|}{Ideal superlattice} & \multicolumn{3}{c}{After optimization} \\ 
		\multicolumn{2}{c|}{\ } & \multicolumn{3}{c|}{(before optimization)} & \multicolumn{3}{c}{\ } \\ \hline 
		Atom & site & $x$ & $y$ & $z$ & $x$ & $y$ & $z$ \\ \hline	
		Ti(1)	& 1a & 0 & 0 & 0 & 0 & 0 & 0 \\
		Ti(2)	& 6g & 1/3 & 0 & 0 & 0.3362 & 0 & 0 \\
		Ti(3)	& 2d & 2/3 & 1/3 & 0 & 2/3 & 1/3 & 0.000455 \\
		Se(1)	& 6i & 1/9 & 2/9 & 0.2573 & 0.1134 & 0.2267 & 0.2561 \\
		Se(2)   & 6i & 4/9 & 2/9 & 0.2573 & 0.4447 & 0.2224 & 0.2548 \\
		Se(3)	& 6i & 4/9 & 8/9 & 0.2573 & 0.4446 & 0.8891 & 0.2572 \\
		Cu	& 1b & 0 & 0 & 1/2 & 0 & 0 & 1/2 \\		
	\end{tabular}
	\end{ruledtabular}
\caption{\label{table:crystal_structure_cu}Comparison of the crystal structure of Cu(TiSe$_2$)$_9$ before optimization, as derived from the stucture of TiSe$_2$, with the relaxed structure obtained by minimizing the forces on the atoms.  The space group of the crystal is P-3m1.}
\end{table}

\begin{figure*}
   \includegraphics[width=\textwidth]{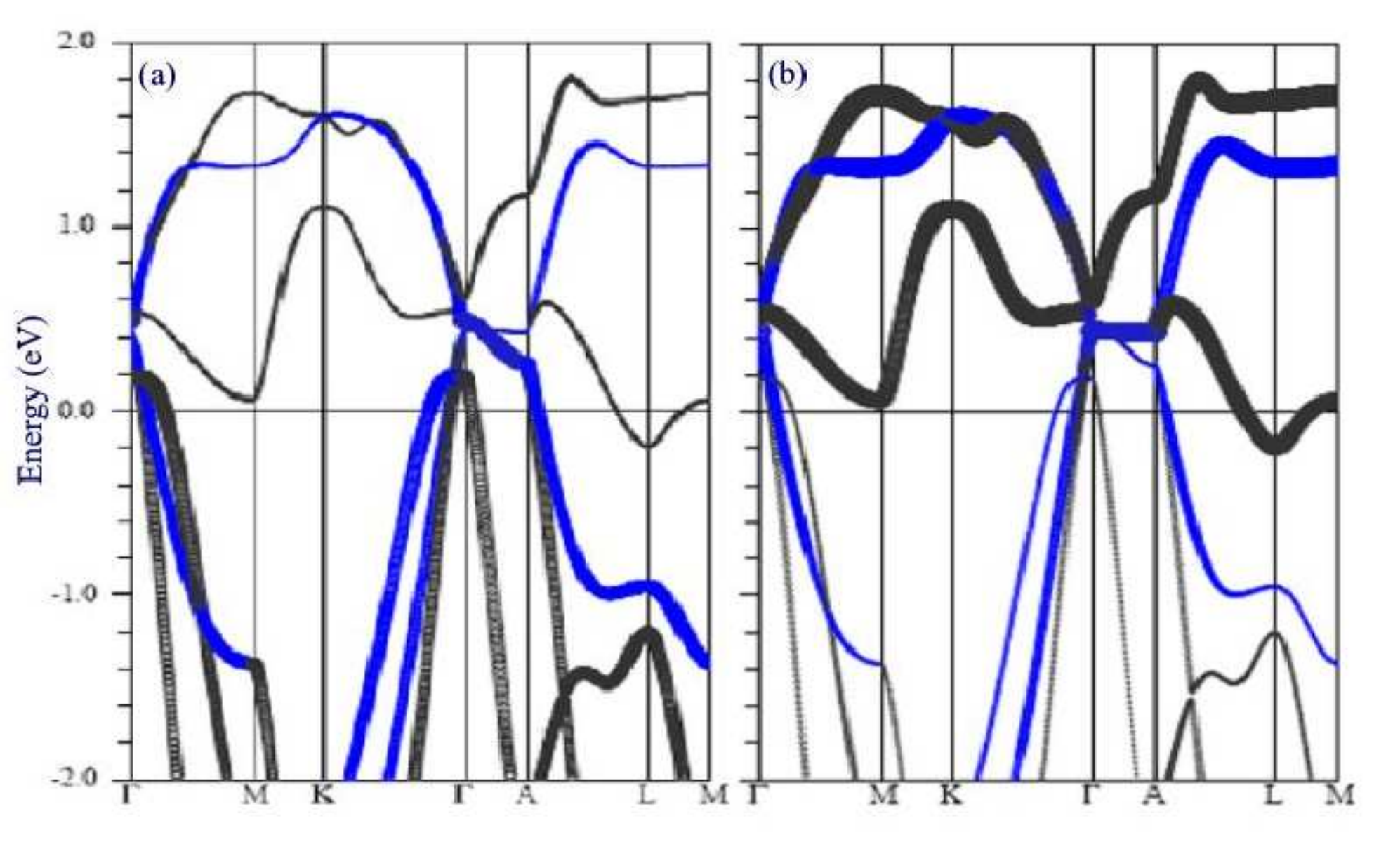}
   \caption{\label{fig:energybands1}(color online)Band character plot for TiSe$_2$ along high symmetry directions in the Brillouin zone.  In (a), the contribution from Se 4p states is shown.  The width of the band lines is proportional to the contribution of the Se 4p states to these bands.  In (b), the contribution of the Ti 3d states to the energy bands is emphasized. The Fermi level is at zero energy.}
\end{figure*}

\begin{figure*}
   \includegraphics[width=\textwidth]{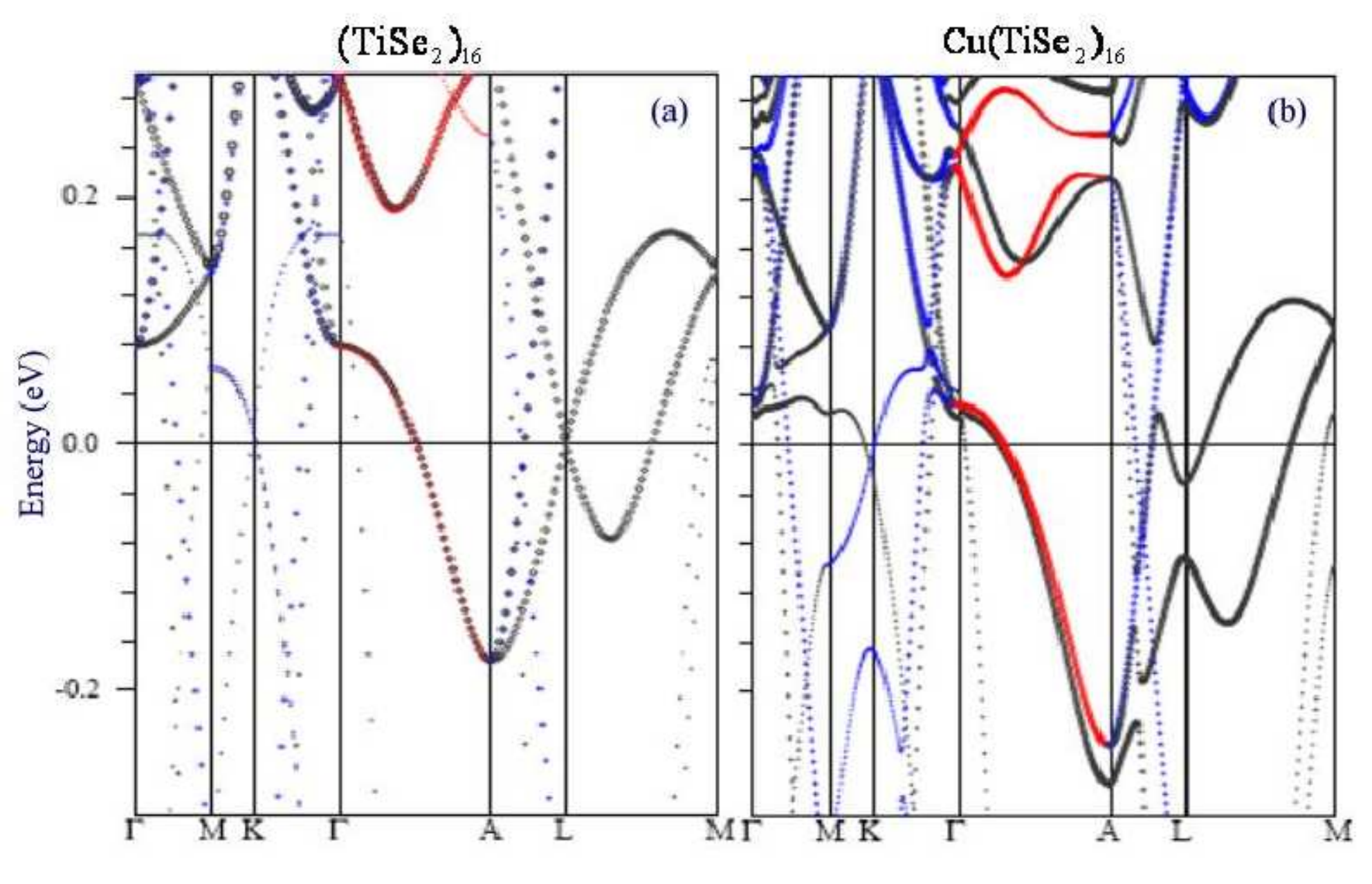}
   \caption{\label{fig:energybands4x4}(color online)Energy bands for the 4x4x1 TiSe$_2$ supperlattice are shown in (a), while those for Cu(TiSe$_2$)$_{16}$ are shown in (b).  The contribution of Ti 3d states is emphasized in both plots, the width of the band lines being proportional to this contribution.  The Fermi level is at zero energy.}
\end{figure*}

\begin{figure*}
   \includegraphics[width=\textwidth]{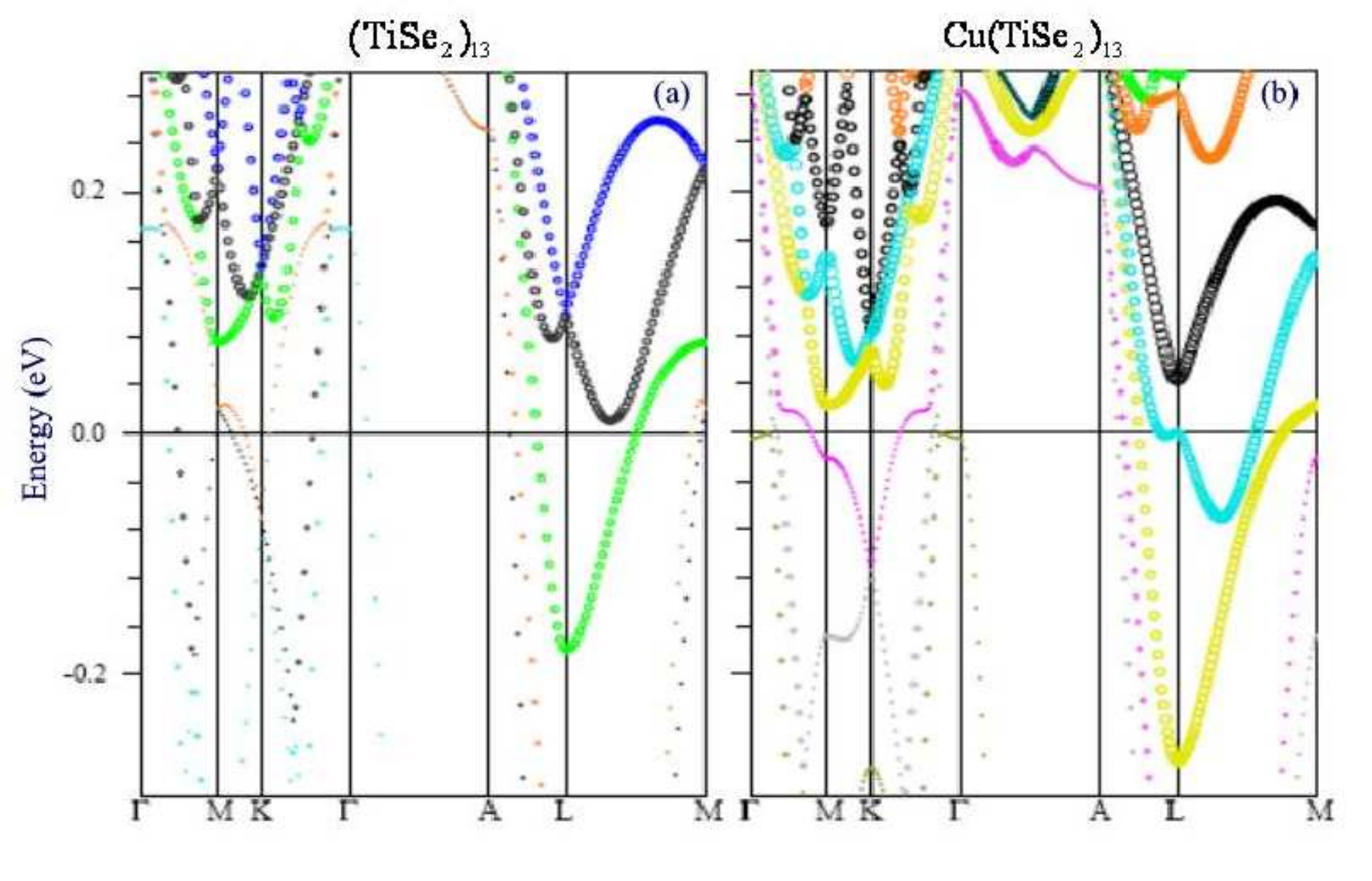}
   \caption{\label{fig:energybandss13xs13}(color online)Energy bands for the $\sqrt{13}$x$\sqrt{13}$x1 TiSe$_2$ superlattice obtained by zone folding of the TiSe$_2$ bands, and the energy bands for Cu(TiSe$_2$)$_{13}$, are shown in (a) and (b), respectively.  The Fermi level is at zero energy.  The Ti 3d contribution to each band is proportional to the width of the band line.  Note that while the Fermi level rises by only 0.06 eV, as a result of Cu intercalation, the hole band, with a maximum at 0.17 eV at $\Gamma$ in (a), shifts down in energy to just below Fermi level.}
\end{figure*}

\begin{figure*}
   \includegraphics[width=\textwidth]{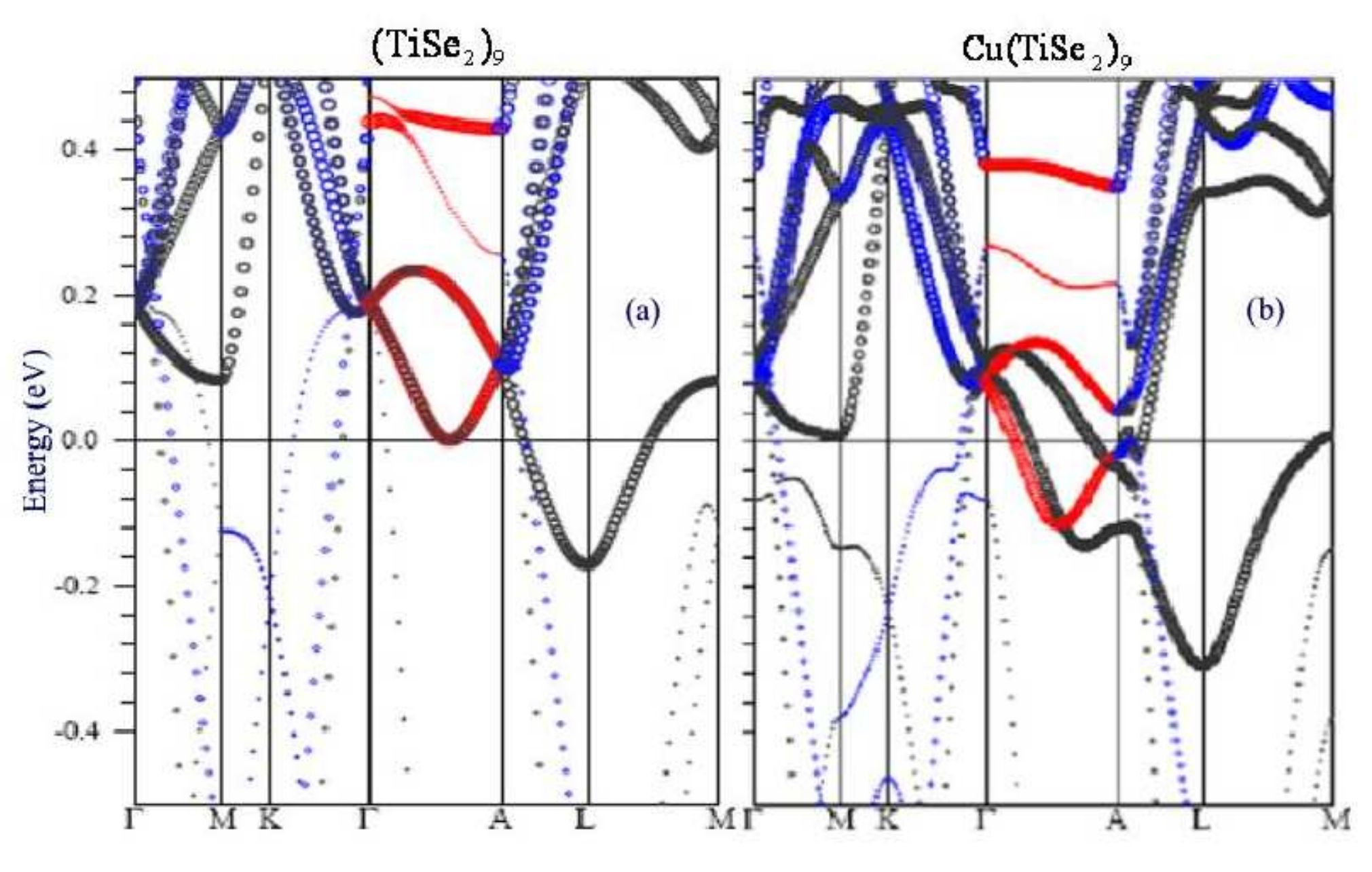}
   \caption{\label{fig:energybands3x3}(color online)Energy bands for the 3x3x1 TiSe$_2$ superlattice, obtained by zone folding of the TiSe$_2$ bands, are shown in (a), while the bands of the copper intercalated compound is shown in (b).  The Fermi level is at zero energy.  The contribution of the Ti 3d states is emphasized.  Note the rise, by 0.08 eV, of the Fermi level in (b), as compared to (a).  The hole band, with a maximum at 0.17 eV at $\Gamma$ in (a), shifts down in energy to -0.08 eV, as a result of copper intercalation.  A similar fate meets the hole band at A.}
\end{figure*}


\section{\label{sec:results_and_discussion}Results and Discussion}
It is well known that the calculation method followed in this work (GGA, PBE) overestimates the unit cell volume and lattice constants and it gives a very large $c/a$ value in layered compounds~\cite{Wu_Cohen:2004}.  Indeed, upon optimization, Titov et al.~\cite{Titov_Kuranov:2001} obtained a value for $c/a$ in TiSe$_2$ which is about 5\% larger than the experimental value.  Since the lattice constants are known for the systems we study here, we perform the calculations using these experimental values, but we allow the internal coordinates to relax.

The unit cell of TiSe$_2$, shown in Fig.~\ref{fig:crystal}a, consists of one Ti atom and two Se atoms.  The crystal coordinates of these atoms are given by Ti(0,~0,~0), Se1(1/3,~2/3,~$z$), and Se2(2/3,~1/3,~$-z$).  Upon minimizing the forces on each atom to below 2 mRy/Bohr, it is found that $z$=0.2573; the distance between the Ti plane and an adjacent Se plane is thus 1.546\AA.  The Brillouin zone (BZ) corresponding to the TiSe$_2$ structure is shown in Fig.~\ref{fig:crystal}b, and the calculated energy bands along some high symmetry directions in the BZ are shown in Fig.~\ref{fig:energybands1}.  In Fig.~\ref{fig:energybands1}a the contribution to the bands from the Se 4p states is shown, while in Fig.~\ref{fig:energybands1}b the contribution from the Ti 3d states is displayed.  The bands are in broad agreement with previous calculations~\cite{Zunger_Freeman:1978}.  They are also in good agreement with recent ARPES study~\cite{Cercellier_Monney:2007} which shows that 1T-TiSe$_2$ has a semimetallic band structure with a negative energy gap E$_\text{G}$ = -70$\pm$15 meV, the Se 4p band having a maximum at 30$\pm$10 meV and the Ti 3d band having a minimum at -40$\pm$5 meV with respect to the Fermi energy.  By examining these plots, it is observed that the electron pocket at L is derived mainly from Ti 3d states, while the hole pockets at $\Gamma$ and A are derived from Se 4p states with some significant hybridization with the Ti 3d states.

Upon intercalation by copper, the $c$-axis is increased slightly, and some slight rearrangement of the positions of the Se atoms takes place.  The new positions are found by relaxing the structure so as to minimize the forces acting on the atoms.  The lattice constants are held fixed at the experimental values, while the internal coordinates are allowed to change.  We find that the incorporation of copper atoms between TiSe$_2$ layers leads to only a slight modification in the atomic positions.  To illustrate this point, we present in Table~\ref{table:crystal_structure_cu} the coordinates of the inequivalent atoms for the case of Cu(TiSe$_2$)$_9$ in the ideal superlattice structure (before optimization) and the corresponding coordinates after optimization.  In the three compounds, Cu$_x$TiSe$_2$, for $x=1/16$, $1/13$, and $1/9$, the Se atoms are displaced in the $z$-direction by $\delta z$ where -0.006\AA\ $<$ $\delta z$ $<$ 0.009\AA, whereas they are displaced in the $a$-$b$ plane by less than 0.037\AA.  Because of the slight displacement in the $z$-direction of the Se atoms, they no longer all lie in a single plane as is the case in undoped TiSe$_2$.  Nevertheless, our results indicate that the disturbance caused by intercalating copper into the van der Waal's gaps of TiSe$_2$ is small as far as its effect on the atomic positions is concerned.  This is not unusual, for a similar situation holds in graphite intercalation compounds where, besides an increase in the c-lattice constant, the effect on the structure of the carbon planes is minimal.  This is a consequence of the fact that, both in TiSe$_2$ and graphite, the bonding within each layer ( in TiSe$_2$ a layer consists of three atomic sheets) is of the strong covalent type.

The intercalation of copper into the van der Waal's gap in TiSe$_2$ is stabilized by charge transfer from copper to the TMDC layers.  Here we calculate the charge on each atom  using Bader's quantum theory of ``atoms in molecules'' (AIM)~\cite{Bader:1990}.  In this theory space is divided among the atoms so that each atom occupies a definite volume bounded by a surface at which the flux in the gradient of the electron density $\rho(\vec r)$ vanishes,
\[
	\hat{n}(\vec{r}) \cdot \vec{\nabla}\rho(\vec{r}) =0
\]
where $\hat{n}(\vec{r})$ is a unit vector normal to this bounding surface at $\vec r$.  The number of electrons in a given atom, occupying a volume $\Omega$, is then given by
\[
N_e = \int_\Omega \rho(\vec r) d^3r.
\]
For undoped TiSe$_2$ we find that Ti is positively charged with a charge of 1.575$e$ while each Se has a negative charge of -0.787$e$.  In the intercalated compounds Cu(TiSe$_2$)$_x$ we find that the copper atom acquires a positive charge of 0.445$e$, 0.444$e$, and 0.450$e$, for $x$ = 9, 13, and 16, respectively.  To illustrate how the charge is distributed we present in Table~\ref{table:charge_of_atom} the charge on each atom in the case of Cu(TiSe$_2$)$_9$, as calculated by Bader's AIM theory.

\begin{table}[b]
  \begin{ruledtabular}
  \begin{tabular}{l c c c}
	Atom & Site & $Q/e$ & $\Delta Q/e$ \\ \hline
	Ti(1) & 1a & 1.551 & -0.024 \\ 
  	Ti(2) & 6g & 1.563 & -0.012 \\
	Ti(3) & 2d & 1.577 & +0.002 \\
	Se(1) & 6i & -0.820 & -0.033 \\
	Se(2) & 6i & -0.801 & -0.014 \\
	Se(3) & 6i & -0.796 & -0.009 \\
	Cu & 1b & 0.445 & +0.445 \\
	\end{tabular}
	\end{ruledtabular}
\caption{\label{table:charge_of_atom}The charge on each inequivalent site in Cu(TiSe$_2$)$_9$, as indicated by AIM theory, and the corresponding change, resulting from copper intercalation, in the charge on each site.  Note that the charge of an electron is -$e$, so that $e$ is positive.}
\end{table}

For the copper intercalated systems, the energy bands appear to be far more complicated than in undoped TiSe$_2$ because the unit cell has many more atoms.  A direct comparison between the undoped TiSe$_2$ energy bands and those of the doped compounds is not very helpful in clearly elucidating the effect of copper intercalation.  Most of the bands in the intercalated compounds result from the zone folding, into the smaller BZ, of the bands in the larger BZ of the pristine compound.  Therefore, in order to gauge accurately the effect of intercalation with copper on the energy bands, it is more useful to compare the bands in the Cu intercalated TiSe$_2$ with those that result from zone folding the pristine TiSe$_2$ bands into the smaller BZ characteristic of the intercalated compounds.  For example, the compound Cu$_{1/16}$TiSe$_2$ has the chemical formula Cu(TiSe$_2$)$_{16}$, and except for a slight variation in the c lattice parameter and an extra Cu atom, its unit cell would be the same as that of a 4x4x1 superlattice of pristine TiSe$_2$, which also consists of 16 TiSe$_2$ units. To understand the effect of the addition of copper in this case, we compare the energy bands of Cu(TiSe$_2$)$_{16}$ with those of (TiSe$_2$)$_{16}$, the 4x4x1 undoped TiSe$_2$ superlattice.  Similar arguments apply in the case of intercalation compounds for other values of $x$, the copper concentration.

The energy bands of the systems studied in this work are shown in Figs.~\ref{fig:energybands4x4},~\ref{fig:energybandss13xs13}, and~\ref{fig:energybands3x3}.  In Fig.~\ref{fig:energybands4x4} we compare the energy bands of the 4x4x1 superlattice with those of Cu(TiSe$_2$)$_{16}$.  It should be noted that for the 4x4x1 superlattice, the BZ is 16 times smaller than that for pristine TiSe$_2$.  The M and L points in the pristine TiSe$_2$ BZ are zone-folded into the $\Gamma$ and A points, respectively, in the superlattice BZ; this explains why features in the band structure at the M and L points in Fig.~\ref{fig:energybands1} now appear at points $\Gamma$ and A, respectively, in Fig.~\ref{fig:energybands4x4}.  For example, in Fig.~\ref{fig:energybands4x4}a, the band along $\Gamma$A, which crosses the Fermi level, and which is derived from Ti 3d states, corresponds to the band along ML in the pristine TiSe$_2$ BZ; the M point is folded to $\Gamma$ and the L point is folded to A.  The effect of intercalation with Cu is seen to be a rise in the Fermi level due to charge transfer from Cu.  The bands derived from Cu states are more than 1 eV below the Fermi level; thus the main influence of Cu intercalation on the electronic properties is a charge transfer to the host layers, since in the immediate vicinity of the Fermi surface, all the bands are derived from Ti and Se states.  For the case of the 3x3x1 and $\sqrt{13}$x$\sqrt{13}$x1 superlattices, the M and L points in the bigger BZ, corresponding to the case when the unit cell contains only one TiSe$_2$ unit, are folded, respectively, into the M an L points in the smaller BZ of the superlattices.

By closely examinning the energy bands in Figs.~\ref{fig:energybands4x4} -~\ref{fig:energybands3x3}, we conclude that the charge transfer from Cu to the host layers leads to a slight rise in the Fermi level, but the host bands derived from Ti 3d states are largely unaffected; the rigid band model is a very good approxiamtion in describing the effect of Cu intercalation in TiSe$_2$ as far as the bands derived from Ti 3d states are concerned.  For the Se 4p derived hole pockets at $\Gamma$ and A points, however, the effect is not described by the rigid band model.  The slight displacement of the Se atoms as a result of intercalation, and the charge transfer from Cu, combine to cause a suppression of the hole band.  For example, for copper concentration $x$=1/9, we see from the energy bands in Fig.~\ref{fig:energybands3x3} that while the Fermi level is raised by $\sim$0.08eV due to Cu intercalation, the hole pockets at $\Gamma$ and A are lowered by $\sim$0.25eV; indeed, these hole pockets are now below the Fermi level, and thus they are completely filled.  Within the exciton condensation mechanism~\cite{Kohn:1967,Halperin_Rice:1968} for CDW transition, such a reduction in the number of holes leads to a suppression of the CDW state.

\begin{figure}
\includegraphics[width=0.5\textwidth]{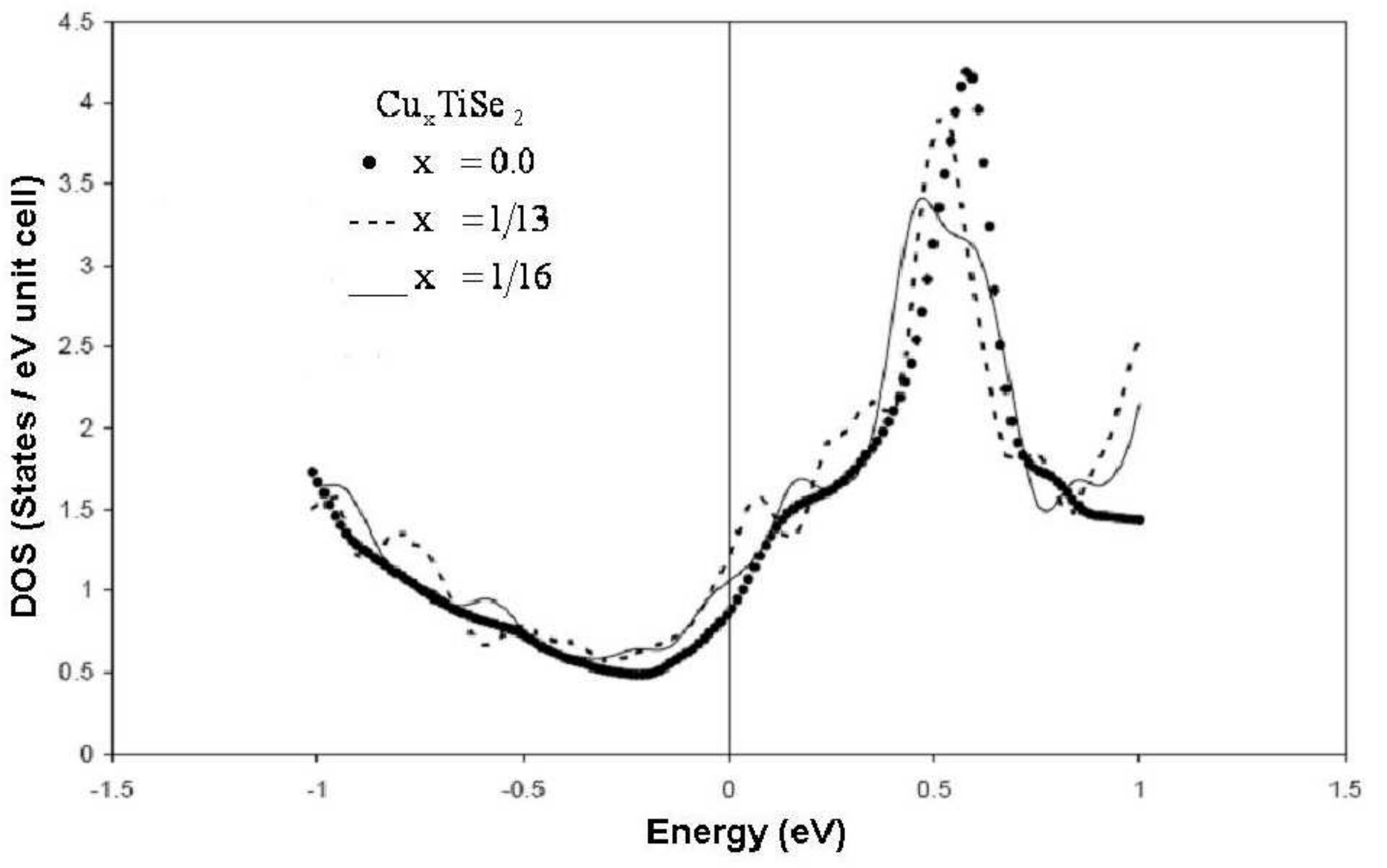}
\caption{\label{fig:density_of_states}Electronic density of states for TiSe$_2$ and two intercalation compounds.  Note the increase in the value of the DOS at the Fermi level, set at the zero of energy, as a result of Cu intercalation.}
\end{figure}

The emergence of superconductivity, as the concentration of Cu increases, is probably related to the suppression of the CDW transition and the concomitant increase in the electronic density of states in the vicinity of the Fermi level.  We calculated the density of states for the various systems considered and in Fig.~\ref{fig:density_of_states} we plot the density of state for pristine TiSe$_2$ and two of the three intercalation compounds considered in this work.  According to the BCS theory of superconductivity,~\cite{Bardeen_Cooper:1957} the transition temperature T$_c$ increases with an increase in the density of states in the immediate vicinity of the Fermi level, the other important factor affecting T$_c$ being the strenght of the electron-phonon coupling.  In Table \ref{table:density_of_states}, we present the calculated density of state at the Fermi energy, in states per eV per one TiSe$_2$ units, for pristine TiSe$_2$ as well as the other various intercalation compounds.  Note the increase in the density of states resulting form intercalation by Cu; the calculated result that this density is largest for the Cu$_{1/13}$TiSe$_2$ case is consistent with the experimental observation that the superconducting transition temperature T$_c$ is largest for Cu concentration given by $x \approx 0.08$.  However, as pointed out earlier, the density of states is not the only factor affecting the value of the transition temperature.

It should be remarked that the density of states at the Fermi energy, N(E$_F$), is not necessarily an increasing function of $x$, the copper concentration, as Table~\ref{table:density_of_states} clearly indicates.  The value of N(E$_F$) is intimately connected to the band structure.  As $x$ increases, the Fermi level rises; consequently, the number of bands crossing the Fermi level may change, or that a band will cross the Fermi level at a new K-point in the BZ where DOS may have a larger or a smaller value than the old K-point.

\begin{table}
	\begin{ruledtabular}
	\begin{tabular}{ l c }
		\textit{Compount} 	& N(E$_F$) \\ \hline
		TiSe$_2$		& 0.86 \\ 
		Cu$_{1/16}$TiSe$_2$	& 1.06 \\ 
		Cu$_{1/13}$TiSe$_2$	& 1.22 \\
		Cu$_{1/9}$TiSe$_2$	& 1.01 \\
	\end{tabular}
	\end{ruledtabular}

\caption{\label{table:density_of_states}The calculated electronic density of state at the Fermi level N(E$_F$), in the states per eV per one TiSe$_2$ unit, for TiSe$_2$ and its copper intercalation compounds.}
\end{table}

\section{\label{conclusion}Conclusion}
In conclusion, we have presented calculations, within the density functional theory, of the electronic energy bands and the density of state of copper intercalated TiSe$_2$.  We have shown that a rigid band model describes reasonably well the effect of the intercalation on Ti 3d derived bands, but not as well for Se 4p derived bands.  As a consequence of charge transfer from the Cu atoms to the host layers, the Fermi level rises, the Se 4p hole pocket gets filled; this is the probable cause for the suppression of the charge density wave transition.  This, coupled with an increase of the electronic density of states at the Fermi level, is perhaps the cause of the emerging superconductivity in these novel intercalation compounds.


\end{document}